\documentclass[runningheads]{llncs}
\usepackage{graphicx}
\usepackage{array}
\newcolumntype{L}{>{\centering\arraybackslash}m{3cm}}
\usepackage{adjustbox}
\usepackage{graphicx}
\usepackage{amsmath}
\usepackage{wrapfig}
\usepackage{hhline}
\usepackage{hyperref}
\usepackage{subfig}
\usepackage{multirow}

\begin{document}
\title{Class Distance Weighted Cross-Entropy Loss for Ulcerative Colitis Severity Estimation}

\titlerunning{Class Distance Weighted Cross-Entropy Loss}

\author{Gorkem Polat\inst{1,2,}\thanks{Corresponding author.}\orcidID{0000-0002-1499-3491}  \and \\
Ilkay Ergenc\inst{3}\orcidID{0000-0003-1539-501X} \and \\
Haluk Tarik Kani\inst{3}\orcidID{0000-0003-0042-9256} \and \\
Yesim Ozen Alahdab\inst{3}\orcidID{0000-0002-1337-9254} \and \\
Ozlen Atug\inst{3}\orcidID{0000-0001-9695-9416} \and \\
Alptekin Temizel\inst{1,2}\orcidID{0000-0001-6082-2573}}

\authorrunning{G. Polat et al.}

\institute{Graduate School of Informatics, Middle East Technical University, Ankara, Turkey 
\and
Neuroscience and Neurotechnology Center of Excellence (NÖROM), Ankara, Turkey
\and
Department of Gastroenterology, Marmara University School of Medicine, Istanbul, Turkey \\
\email{gorkem.polat@metu.edu.tr, ergencilkay@gmail.com, drhtkani@gmail.com, yesimalahdab@yahoo.com, ozlenatug@hotmail.com, atemizel@metu.edu.tr}
}

\maketitle    
\begin{abstract}

In scoring systems used to measure the endoscopic activity of ulcerative colitis, such as Mayo endoscopic score or Ulcerative Colitis Endoscopic Index Severity, levels increase with severity of the disease activity. Such relative ranking among the scores makes it an ordinal regression problem. On the other hand, most studies use categorical cross-entropy loss function to train deep learning models, which is not optimal for the ordinal regression problem. In this study, we propose a novel loss function, class distance weighted cross-entropy (CDW-CE), that respects the order of the classes and takes the distance of the classes into account in calculation of the cost. Experimental evaluations show that models trained with CDW-CE outperform the models trained with conventional categorical cross-entropy and other commonly used loss functions which are designed for the ordinal regression problems. In addition, the class activation maps of models trained with CDW-CE loss are more class-discriminative and they are found to be more reasonable by the domain experts.

\end{abstract}

\section{Introduction}

Deep learning (DL) methods are widely used in the field of gastrointestinal endoscopy for problems such as detection of polyps, artifacts, Barrett's esophagus, and cancer analysis \cite{polat2020endoscopic,ali2021deep,polat2021polyp,ali2022assessing,du2019review}. In particular, recent studies have reported successful results for the estimation of the endoscopic activity of ulcerative colitis (UC) from colonoscopy images  \cite{limdi2021automated,takenaka2022artificial}. UC is a chronic condition caused by persistent inflammation of the colon mucosa and accurate assessment of the disease severity plays a key role in monitoring and treating the disease. However, there are substantial intra- and inter-observer variability in the grading of endoscopic severity \cite{osada2010comparison} and use of computer-aided diagnosis of the UC can eliminate subjectivity and help experts in the monitoring process. On the other hand, more work, such as validation on external datasets and providing better explainability, are needed to increase their adoption in clinics \cite{limdi2021automated}. 

Scoring systems for UC, such as Mayo endoscopic score (MES) \cite{schroeder1987coated} or Ulcerative Colitis Endoscopic Index of Severity (UCEIS) \cite{travis2013reliability}, have several levels (0-3 for MES and 0-8 for UCEIS), which increase in relation to the severity of the disease. Since there is a ranking between the class scores, this problem can be handled as an ordinal regression (or ordinal classification) problem. Although there are many studies on UC endoscopic activity estimation, only a few of these exploit ordinality information. In this study, we propose a novel non-parametric loss function, which respects the ordinal nature of the problem and calculates the cost accordingly.

The main contributions of this paper are as follows:
\begin{enumerate}
\item A new loss function called Class Distance Weighted Cross-Entropy (CDW-CE) is proposed, which can be used in training convolutional neural networks (CNN) estimating the endoscopic severity of UC.  

\item Three separate CNN architectures are trained using cross-entropy, CORN framework \cite{shi2021deep}, cross-entropy with an ordinal loss term (CO2) \cite{albuquerque2021ordinal}, ordinal entropy loss (HO2) \cite{albuquerque2021ordinal}, and CDW-CE. These networks are used to comparatively evaluate the effect of these particular loss functions in estimation of endoscopic severity of UC.

\item We demonstrate through Class Activation Map (CAM) visualizations that models trained with CDW-CE are more class-discriminative and provide better explainability, which are key factors in adoption of computer-aided diagnosis systems for clinical use.
\end{enumerate}

\section{Related Work}

There has been increasing interest in automatically estimating the UC severity from colonoscopy images. Alammari et al. \cite{alammari2017classification} proposed a 9-layer simple CNN architecture to classify frames in colonoscopy videos. They reported that the model can process the 128$\times$128 pixel images in real-time with 67.7\% test set accuracy. Stidham et al. \cite{stidham2019performance} performed one of the earliest comprehensive studies on a large dataset and employed an advanced CNN architecture Inception-v3 \cite{szegedy2016rethinking} to classify images according to MES. Ozawa et al. \cite{ozawa2019novel} used GoogLeNet \cite{szegedy2015going} to classify images into three MES levels (Mayo 0, Mayo 1, and Mayo 2-3) due to lack of severe cases. Takenaka et al. \cite{takenaka2020development} used Inception-v3 \cite{szegedy2016rethinking} to estimate endoscopic remission, histologic remission, and UCEIS score using one of the largest datasets used in studies. Bhambhvani et al. \cite{bhambhvani2021deep} used ResNext-101 model on publicly available HyperKvasir dataset. Yao et al. \cite{yao2021fully} developed a fully automated system that can estimate MES score for the colonoscopy video. Kani et al. \cite{kani2021p099} employed ResNet18 model to classify MES, severe mucosal disease diagnosis, and remission. Gottlieb et al. \cite{gottlieb2021central} estimated the MES and UCEIS for full-length endoscopy videos where the annotation is only provided for the video itself rather than the individual frames. Schwab et al. \cite{schwab2021automatic} used a multi-instance learning approach with ordinal regression methods to estimate UC severity from both frame-level and video-level MES labels. Becker et al. \cite{gutierrez2021training} proposed an end-to-end fully automated system to estimate MES from raw colonoscopy videos directly. They employed a quality checking model to extract readable frames and weak MES labels obtained by the colon-segment-wise scores were assigned to them to train the UC grading model. Different to the previous approaches employing the existing DL models, Luo et al. \cite{luo2022diagnosis}, proposed a new architecture called UC-DenseNet which combines CNN, RNN, and attention mechanisms. Sutton et al. \cite{sutton2022artificial} compared many state-of-the-art CNN models on HyperKVasir \cite{borgli2020hyperkvasir} dataset and reported that DenseNet121 \cite{huang2017densely} outperformed the other models.

Ordinal categories are common in many real-world prediction problems, especially in the healthcare domain. Several loss functions have been introduced recently to use in conjunction with CNNs. Niu et al. \cite{niu2016ordinal} transformed an ordinal regression problem into a series of binary classification sub-tasks based on the work of Li et al. \cite{NIPS2006_019f8b94}. They applied this approach to age estimation from face images and reported better results compared to other ordinal regression approaches such as metric learning and widely used cross-entropy loss function. Although the proposed method provided better results, there were rank inconsistencies in the output classification subtasks. Cao et al. \cite{cao2020rank} proposed a consistent rank logits (CORAL) framework for rank-consistencies by weight sharing in the penultimate layer. They reported that the CORAL framework provided both rank consistency and superior results compared to the previous approaches. Shi et al. \cite{shi2021deep} proposed Conditional Ordinal Regression for Neural Network (CORN) framework to relax the constraint on the penultimate layer of the CORAL framework to increase neural network's capacity by introducing conditional probabilities. The authors reported that the CORN approach performs better than previous methods. A major disadvantage of CORN-like approaches is that they require a change in the model architecture (output layer) and labeling structure. Another approach for ordinal regression problems is to integrate unimodality in the loss function \cite{belharbi2019unimoconstraints,albuquerque2021ordinal}. This approach enforces unimodality by punishing inconsistencies in the posterior probability distribution among adjacent labels. The punishing term is generally added next to the main loss function, where cross-entropy is used mostly. Albuquerque et al. \cite{albuquerque2021ordinal} employed a unimodality approach for the cervical cancer classification by using cross-entropy and entropy losses as main loss functions and reported better performance results compared to other approaches. Through the manuscript, cross-entropy and entropy loss with unimodality loss terms will be referred to as CO2 and HO2 respectively as in \cite{albuquerque2021ordinal}. Another class of the methods is to use regression to predict a single continuous value at the output or using sigmoid activation function on top of it to limit prediction in [0, 1], then using thresholds or probability distributions to convert the output into discrete levels \cite{beckham2016simple,beckham2017unimodal}. However, regression-based approaches assume fixed distances between classes and encoding specific parametric distributions (e.g., Gaussian, Poisson) at the network output restricts the model and prevents scaling to a large number of classes \cite{beckham2017unimodal}. Moreover, tuning parameters in parametric distributions presents another challenge. Regression-based approaches or methods enforcing parametric distributions have been shown to be inferior to other methods in many studies \cite{belharbi2019unimoconstraints,albuquerque2021ordinal}.

Among the studies in the literature, only Schwab et al. \cite{schwab2021automatic} employed two ordinal regression approaches. In their first approach, they applied a CORN-like framework by transforming the output layer into multiple binary subtasks. In their second approach, their models output a continuous value between 0 and 3, and classes are assigned according to the thresholds; however, optimum class thresholds are determined using a search on the dataset, which limits the generalizability of the proposed method. Furthermore, it is not trivial to derive a confidence value for the assessment due to the numeric value, and the method is not compatible with the CAM visualization techniques as it has a single node at the output layer which is responsible for all classes.

In this study, we propose a novel non-parametric loss function called CDW-CE. CDW-CE can be used in conjunction with any model and does not require any changes in the model architecture or labeling structure. Moreover, it does not require setting any thresholds or enforcing a probability distribution and is compatible with CAM visualization techniques.

\section{Class Distance Weighted Cross-Entropy}
\subsection{Motivation}

Cross-entropy loss function, which is widely used in classification tasks, does not take into account how probabilities of the predictions are distributed among the non-true classes (Equation \ref{eqn:CE}):


\begin{equation} \label{eqn:CE}
    \textrm{\textbf{CE}}=-\sum_{i=0}^{N-1}{y_i \times \log{\hat{y}_i}}=-\log{\hat{y}_{c}}
\end{equation}

\noindent where $i$ is the index of the class in the output layer, $c$ is the index of ground-truth class, $y$ is the ground-truth label, and $\hat{y}$ is the prediction. Since one-hot encoding is used for the ground-truth labels of the classes at the output layer, $y_i=0$ $\forall i\neq c$. Eventually, cross-entropy loss only evaluates the predicted confidence of the true class. However, when there is a ranking among the output classes, class mispredictions become important, too. For example, in an ordinal class structure from 0 to 9, predicting 0 for class 9 is much worse than predicting 8. A better loss function would evaluate this ranking and penalize more if the predictions are away from the true class (see Table \ref{tab:ce_loss_samples}). Since the predictions farther from the correct classes are not penalized more than the closer classes, cross-entropy is not an optimum loss function for the ordinal classes.

\begin{table}[]
\vspace{-5mm}
\caption{Three sample cases that result in the same cross-entropy loss where Class 0 is the true class. Assuming that the classes have an ordinal relation, a more suitable loss function should favor Case 1 by assigning the lowest cost and Case 3 should have the highest cost. }
\label{tab:ce_loss_samples}
\begin{adjustbox}{center}
\begin{tabular}{c|ccc}
\textbf{Classes} & \textbf{Case 1} & \textbf{Case 2} & \textbf{Case 3} \\ \hline
0                & 0.6               & 0.6               & 0.6               \\
1                & 0.3               & 0.1               & 0                 \\
2                & 0.1               & 0.3               & 0.1               \\
3                & 0                 & 0                 & 0.3              
\end{tabular}
\end{adjustbox}
\vspace{-5mm}
\end{table}

\subsection{Class Distance Weighted Cross-Entropy Loss Function}

We propose a non-parametric loss function CDW-CE that evaluates the confidences of non-true classes instead of the true class confidence as in cross-entropy (Equation \ref{eqn:CDW-CE}). Firstly, we penalize how much each misprediction deviates from the true value using log loss. Since one-hot encoding is used for encoding the class labels for multi-class classification problems, predicted confidences for the non-true classes should be equal to zero. Secondly, we introduce a coefficient for the loss of each class, which utilizes the distance to the ground-truth class and increases in relation to that distance.

\begin{equation} \label{eqn:CDW-CE}
    \textrm{\textbf{CDW-CE}} = -\sum_{i=0}^{N-1} {\log(1-\hat{y}_i) \times |i-c|^{\alpha}}
\end{equation}

\noindent where $c$ is the index of the ground-truth class and power term $\alpha$ is a hyperparameter that determines the strength of the coefficient. Eventually, the logarithmic function inside the summation is calculated for every non-true class.

\section{Experiments}

\subsection{Dataset}

LIMUC dataset \cite{gorkem_polat_2022_5827695}, a publicly available UC dataset labeled according to the MES, was used to train CNN models that employ different loss functions. There are 11276 images from 564 patients in the LIMUC dataset and all images have been reviewed and annotated by at least two expert gastroenterologists. All images have a size of 352$\times$288 and Mayo score distribution is as follows: 6105 (54.14\%) Mayo 0, 3052 (27.7\%) Mayo 1, 1254 (11.12\%) Mayo 2, and 865 (7.67\%) Mayo 3. 15\% of the images (1686 images from 85 patients) have been used as the test set and the rest (9590 images from 479 patients) for the 10-fold cross-validation by forming train-validation set pairs. All splittings have been performed at the patient-level, randomly, and preserving class ratios. 

\subsection{Training details}
Three commonly used CNN architectures, ResNet18 \cite{he2016deep}, Inception-v3 \cite{szegedy2016rethinking}, and MobileNet-v3-large \cite{howard2019searching} have been trained with different loss functions. ResNet and Inception model families are commonly used architectures for UC severity estimation \cite{stidham2019performance,ozawa2019novel,takenaka2020development,bhambhvani2021deep,gutierrez2021training,schwab2021automatic,yao2021fully}. MobileNet-v3-large is a more recent model that stands out with its speed and performance, making it a suitable choice for real-time UC severity estimation from video frames. Random rotation ($0^{\circ}-360^{\circ}$) and horizontal flipping were used as data augmentation and weights were initialized from pretrained models on IMAGENET dataset \cite{russakovsky2015imagenet}. Adam optimizer with a learning rate of $2e-4$ and learning rate scheduling with a scaling factor of 0.2 was applied if there were no increase in the validation set accuracy for the last 10 epochs. Early stopping was used to terminate training when performance did not increase in the last 25 epochs. The best model checkpoint on the validation set of each fold is used for the performance measurement on the test set. PyTorch framework \cite{paszke2019pytorch} were used for the implementation of the study and CNN models were adapted from TorchVision package.

The proposed model has been evaluated against three state-of-the-art approaches specifically designed for the ordinal regression tasks: CORN framework, CO2, and HO2 and cross-entropy (CE) loss function is used as the main baseline. For the training of CO2 and HO2 models, main loss function (either cross-entropy or entropy loss) is scaled with a $\lambda$ coefficient as in original paper implementation. Hyperparameter tuning for the $\lambda$ were performed using values in \{0.1, 0.01, 0.001\} by performing 10-fold cross validation.

\begin{table}[t!]
\caption{Experiment results for all Mayo scores.}
\label{tab:experiment_results_mayo_scores}
\begin{adjustbox}{center}
\begin{tabular}{cllccc} 
\hline
                          & \multicolumn{1}{c}{} & \multicolumn{1}{c}{\textbf{Loss Function}} & \textbf{ResNet18}       & \textbf{Inception-v3}   & \textbf{MobileNet-v3-Large}  \\ 
\hline
\multirow{5}{*}{QWK}      &                      & Cross-Entropy                              & 0.8296 ± 0.014          & 0.8360 ± 0.011          & 0.8302 ± 0.015               \\
                          &                      & CORN                                       & 0.8366 ± 0.007          & 0.8431 ± 0.009          & 0.8412 ± 0.010               \\
                          &                      & CO2                                        & 0.8394 ± 0.009          & 0.8482 ± 0.009          & 0.8354 ± 0.009               \\
                          &                      & HO2                                        & 0.8446 ± 0.007          & 0.8458 ± 0.010          & 0.8378 ± 0.007               \\
                          &                      & CDW-CE                                     & 0\textbf{.8568 ± 0.010} & \textbf{0.8678 ± 0.006} & \textbf{0.8588 ± 0.006}      \\ 
\hline
\multirow{5}{*}{F1}       &                      & Cross-Entropy                              & 0.6720 ± 0.026          & 0.6829 ± 0.023          & 0.6668 ± 0.028               \\
                          &                      & CORN                                       & 0.6809 ± 0.014          & 0.6832 ± 0.013          & 0.6847 ± 0.020               \\
                          &                      & CO2                                        & 0.6782 ± 0.014          & 0.6846 ± 0.016          & 0.6793 ± 0.012               \\
                          &                      & HO2                                        & 0.6856 ± 0.016          & 0.6901 ± 0.008          & 0.6741 ± 0.030               \\
                          &                      & CDW-CE                                     & \textbf{0.7055 ± 0.021} & \textbf{0.7261 ± 0.015} & \textbf{0.7254 ± 0.010}      \\ 
\hline
\multirow{5}{*}{Accuracy} &                      & Cross-Entropy                              & 0.7566 ± 0.015          & 0.7600 ± 0.012          & 0.7564 ± 0.011               \\
                          &                      & CORN                                       & 0.7591 ± 0.009          & 0.7600 ± 0.008          & 0.7613 ± 0.012               \\
                          &                      & CO2                                        & 0.7601 ± 0.008          & 0.7654 ± 0.008          & 0.7572 ± 0.009               \\
                          &                      & HO2                                        & 0.7625 ± 0.011          & 0.766 ± 0.010           & 0.7583 ± 0.005               \\
                          &                      & CDW-CE                                     & \textbf{0.7740 ± 0.011} & \textbf{0.7880 ± 0.011} & \textbf{0.7759 ± 0.010}      \\ 
\hline
\multirow{5}{*}{MAE~}     &                      & Cross-Entropy                              & 0.2581 ± 0.018          & 0.2526 ± 0.013          & 0.2563 ± 0.012               \\
                          &                      & CORN                                       & 0.2517 ± 0.012          & 0.2497 ± 0.010          & 0.2480 ± 0.012               \\
                          &                      & CO2                                        & 0.2497 ± 0.011          & 0.2404 ± 0.008          & 0.2524 ± 0.010               \\
                          &                      & HO2                                        & 0.2460 ± 0.011          & 0.2424 ± 0.011          & 0.2487 ± 0.005               \\
                          &                      & CDW-CE                                     & \textbf{0.2300 ± 0.011} & \textbf{0.2147 ± 0.010} & \textbf{0.2272 ± 0.011}      \\
\hline
\end{tabular}
\end{adjustbox}
\end{table}

\subsection{Evaluation Metrics}

Quadratic Weighted Kappa (QWK) is used as the main performance metric as it is suitable for both imbalanced and ordinal data. In addition, Mean Absolute Error (MAE), which is a commonly used performance metric in ordinal regression problems, accuracy, and macro F1 metrics are given in Table \ref{tab:experiment_results_mayo_scores}. In addition to the MES prediction, inflammatory bowel disease (IBD) experts are also interested in the estimation of endoscopic remission (Mayo 0 or 1) and moderate to severe disease (Mayo 2 or 3) as defined in the European Medicine Agency and the US Food and Drug Administration guidelines on UC drug development \cite{REINISCH20191673}. Trained CNN models for MES estimation were used for remission classification performance measurements by grouping the related Mayo subscores, without any new training. Cohen's Kappa, F1, and accuracy scores for remission classification are reported in Table \ref{tab:experiment_results_remission}.

\begin{table}[t!]
\caption{Experiment results for remission classification.}
\label{tab:experiment_results_remission}
\begin{adjustbox}{center}
\begin{tabular}{cllccc} 
\hline
                          &  & \textbf{Loss Function} & \textbf{ResNet18}                & \textbf{Inception-v3}            & \textbf{MobileNet-v3-Large}       \\ 
\hline
\multirow{5}{*}{Kappa}    &  & Cross-Entropy          & 0.8077 ± 0.023                   & 0.8074 ± 0.021                   & 0.8122 ± 0.018                    \\
                          &  & CORN                   & 0.8191 ± 0.021                   & 0.8077 ± 0.022                   & 0.8203 ± 0.016                    \\
                          &  & CO2                    & 0.8185 ± 0.020                   & 0.8243 ± 0.011                   & 0.8067 ± 0.020                    \\
                          &  & HO2                    & 0.8318 ± 0.015                   & 0.8251 ± 0.015                   & 0.8283 ± 0.018                    \\
                          &  & CDW-CE                 & \textbf{0.8521~}±\textbf{~0.016} & \textbf{0.8598~}±\textbf{~0.012} & \textbf{0.8592~}±\textbf{~0.012}  \\ 
\hline
\multirow{5}{*}{F1}       &  & Cross-Entropy          & 0.8419 ± 0.018                   & 0.8420 ± 0.017                   & 0.8451 ± 0.016                    \\
                          &  & CORN                   & 0.8511 ± 0.016                   & 0.8425 ± 0.018                   & 0.8523 ± 0.013                    \\
                          &  & CO2                    & 0.8513 ± 0.015                   & 0.8561 ± 0.009                   & 0.8404 ± 0.017                    \\
                          &  & HO2                    & 0.8618 ± 0.012                   & 0.8565 ± 0.011                   & 0.8583 ± 0.015                    \\
                          &  & CDW-CE                 & \textbf{0.8785~}±\textbf{~0.013} & \textbf{0.8847~}±\textbf{~0.010} & \textbf{0.8842~}±\textbf{~0.010}  \\ 
\hline
\multirow{5}{*}{Accuracy} &  & Cross-Entropy          & 0.9436 ± 0.009                   & 0.9432 ± 0.007                   & 0.9456 ± 0.005                    \\
                          &  & CORN                   & 0.9473 ± 0.007                   & 0.9429 ± 0.008                   & 0.9473 ± 0.006                    \\
                          &  & CO2                    & 0.9461 ± 0.008                   & 0.9479 ± 0.004                   & 0.9444 ± 0.006                    \\
                          &  & HO2                    & 0.9507 ± 0.005                   & 0.9485 ± 0.005                   & 0.9504 ± 0.005                    \\
                          &  & CDW-CE                 & \textbf{0.9566~}±\textbf{~0.005} & \textbf{0.9590~}±\textbf{~0.003} & \textbf{0.9588~}±\textbf{~0.005}  \\
\hline
\end{tabular}
\end{adjustbox}
\end{table}

\begin{figure}[t!]
  \centering
  \includegraphics[width=\textwidth,keepaspectratio]{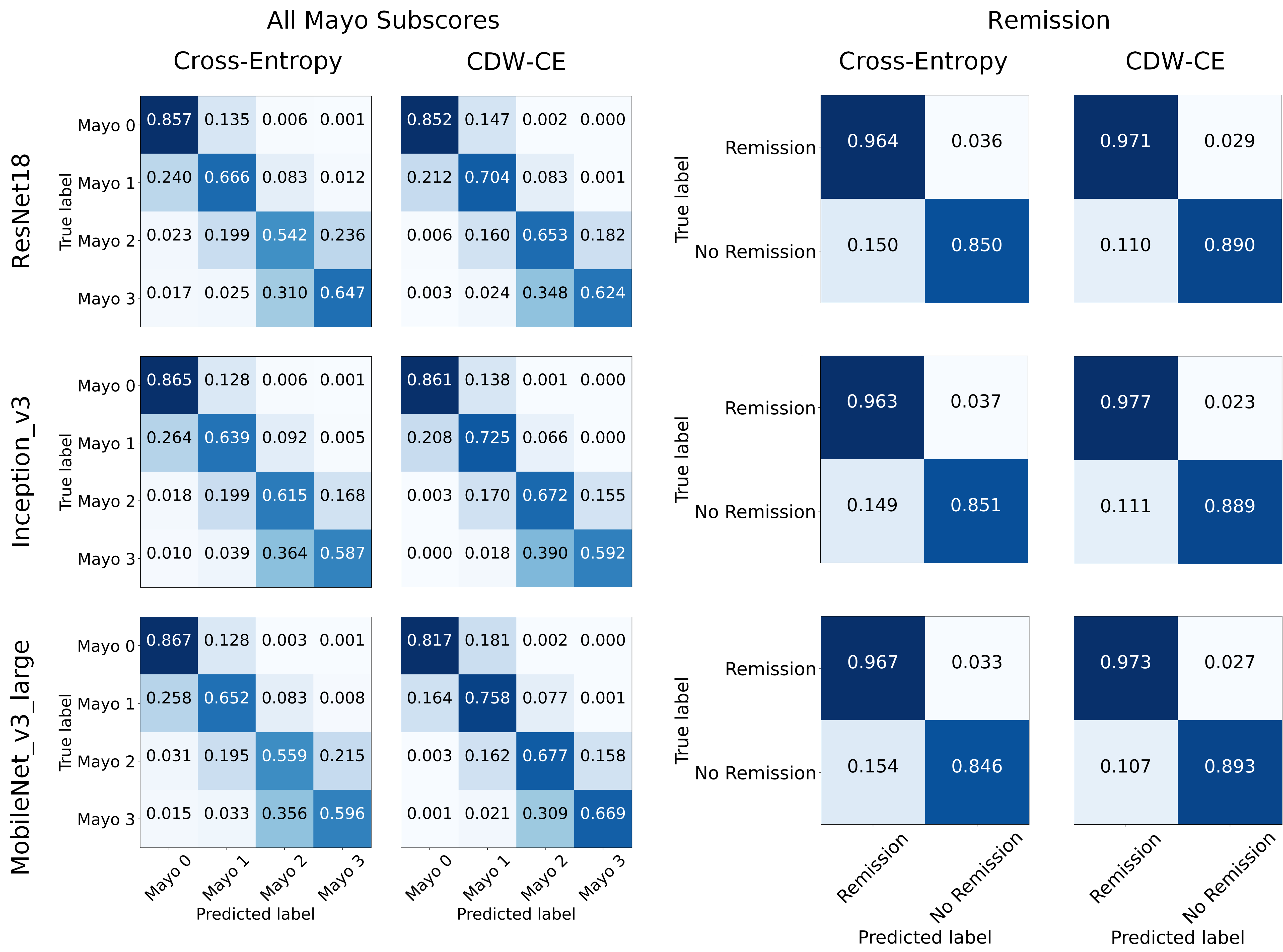}
  \caption{Mean confusion matrix of each CNN model trained with CE and CDW-CE for all Mayo classes and remission classification.}
  \label{fig:models_confusion_matrix}
\end{figure}

Each CNN model has been trained on a different fold and performance measurements were obtained on the initially separated test set; as a result, each architecture has ten different results. Reported performance results in Tables \ref{tab:experiment_results_mayo_scores} and \ref{tab:experiment_results_remission} refer to the average and standard deviation of 10 folds. To observe how much the performance of each class changes  with CDW-CE compared to cross-entropy for three different models, confusion matrices are demonstrated in Figure \ref{fig:models_confusion_matrix} for both all Mayo classes and remission classification. Confusion matrices produced for each fold were normalized across true labels, then, the mean confusion matrix was obtained by getting the average of 10 normalized confusion matrix.

\subsection{Penalization Factor Analysis}

\begin{figure}[t!]
  \centering
  \includegraphics[width=\textwidth]{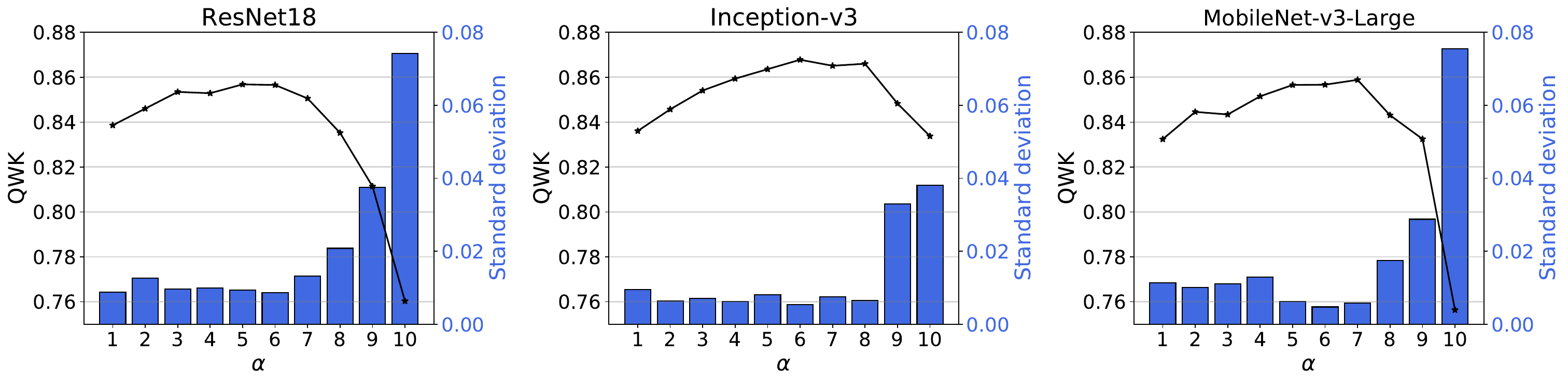}
  \caption{Change of mean and standard deviation of QWK scores according to varying $\alpha$ for three models.}
  \label{fig:power_analysis}
\end{figure}

Power term $\alpha$ in the CDW-CE loss provides a control over to what extent the more distant classes are penalized. As the $\alpha$ increases, the distant classes are penalized more intensely. However, this penalization factor may vary depending on external factors such as the dataset, number of labels, and the employed CNN model. We have analyzed different $\alpha$ values to determine the optimum for each CNN model. The results in Tables \ref{tab:experiment_results_mayo_scores} and \ref{tab:experiment_results_remission} for CDW-CE are the results of the models trained with the experimentally determined optimum $\alpha$. For each CNN model, mean and standard deviation of the QWK scores for varying $\alpha$ are given in Figure \ref{fig:power_analysis}.

\section{Class Activation Maps (CAM)}

\begin{figure}[t!]
  \centering
  \includegraphics[width=\textwidth,height=\textheight,keepaspectratio]{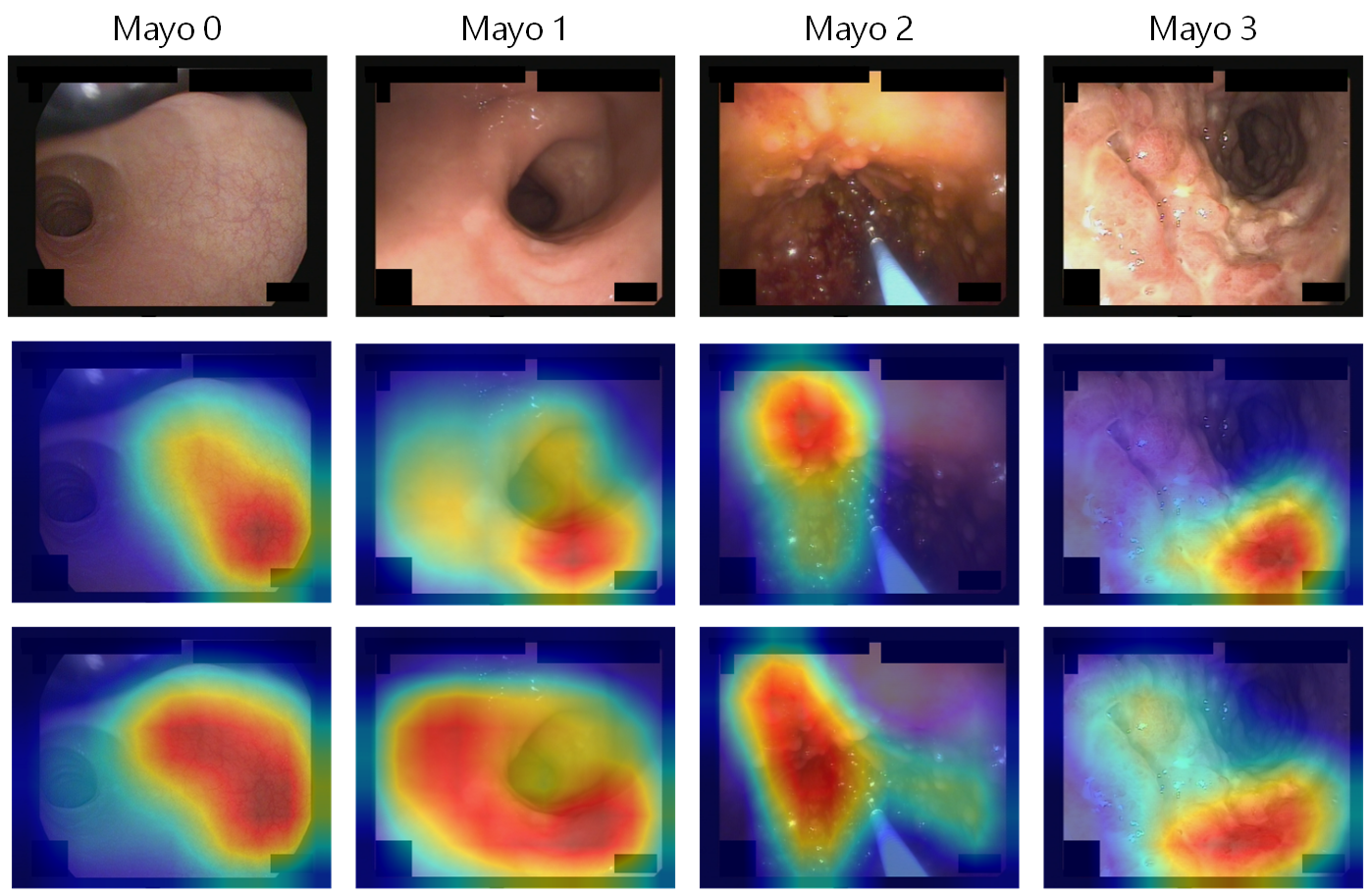}
  \caption{Original images (top row) and their CAM visualizations of the ResNet18 model trained with CE (middle row) and CDW-CE (bottom row) losses. The model trained with CDW-CE highlights broader and more relevant areas related to the disease.}
  \label{fig:cam_examples}
\end{figure}

To make a CNN model's decision more transparent and interpretable, several visualization techniques have been proposed \cite{zhou2016learning,selvaraju2017grad}. CAM visualizations allow observation of the prominent regions used by the models in making their predictions, which is a particularly important aspect in the medical domain. Models which make their predictions using similar regions with the experts would be more likely to be adopted and build more trust with end-users. Such visualizations can also be used as an another comparison criteria and allow assessing different models when their performances are similar (i.e., models with more reasonable activation maps can be chosen instead of others even if their performances are exactly the same).  In addition, it provides a means for developers to debug their approach and check any potential biases in the model's predictions \cite{selvaraju2017grad}. We have generated CAM visualizations using the technique in \cite{zhou2016learning}. Since CAM is produced specifically for each class, it highlights the class-specific discriminative regions only for the target class. In Figure \ref{fig:cam_examples}, two ResNet18 models trained with CE and CDW-CE losses are used to generate CAMs for different images in the test set. Although both models correctly predict the class scores for the given examples, their CAMs differ considerably.   

\begin{figure}[t!]
  \centering
  \includegraphics[width=\textwidth,keepaspectratio]{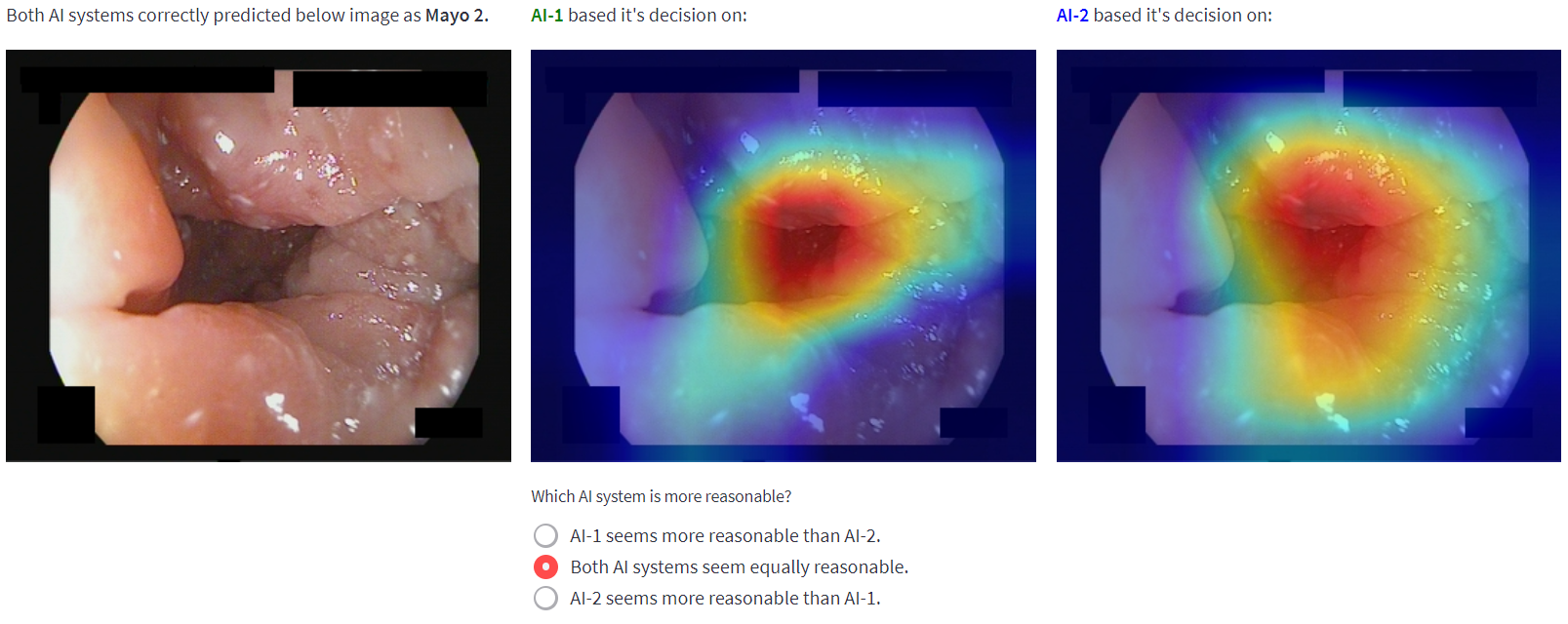}
  \caption{User interface provided to the experts displays the CAM visualizations alongside the image. Experts are asked to evaluate the spots used in decision making process of CE and CDW-CE and choose the one which they think is more reasonable to them (i.e., more aligned with their decision making).} 
  \label{fig:cam_experiment_screenshot}
\end{figure}

To make a quantitative evaluation of CAMs produced by two models trained with different loss functions, we asked three IBD experts to choose which one was more compatible with symptomatic areas in the tissue (i.e., more aligned with the regions they considered in their decision making). We also allowed them to specify that both are equally reasonable, when they are not able to decide between two CAM visualizations. We showed the experts a total of 240 images (60 images from each class), which were correctly predicted by the two models. Only the original image and the two CAM visualizations overlaid onto original images were shown to experts. CAM images produced by the models for each new image were randomly named as AI-1 (Artificial Intelligence 1) and AI-2. Clinicians were asked to make a choice between three options without having the knowledge of model-CAM visualization correspondence (Figure \ref{fig:cam_experiment_screenshot}).  

\section{Results and Discussion}

Table \ref{tab:experiment_results_mayo_scores} shows that CE loss is the worst performing among all models, indicating that this widely used loss function is not optimal and approaches taking ordinality into account are more preferable. Unimodality approaches compare favorably to CORN framework for the ResNet18 and Inception-v3 models and only behind for the MobileNet-v3-large model; however, with an insignificant margin. HO2 results are mostly better than CO2, which is aligned with results reported in the literature \cite{albuquerque2021ordinal}. CDW-CE outperforms other approaches in all experiments. For all models, CDW-CE results refer to the training with the optimum $\alpha$, which are 5, 6, and 7 for the ResNet18, Inception-v3, and MobileNet-v3-large, respectively. Similar performance comparison can also be observed for remission classification in Table \ref{tab:experiment_results_remission}. CO2 and CORN framework have very similar performances. On the other hand, HO2 outperformed CORN framework for all models indicating that it is better at centering estimations around the true class. CDW-CE loss has the highest score for all performance metrics and CNN models. When we observe the individual class performances, Figure \ref{fig:models_confusion_matrix} shows that CDW-CE loss significantly reduces the mispredictions which are in two-class distance or more to the true class. Although sensitivity of edge classes (Mayo 0 and Mayo 3) remained the same or even decreased for some models, intermediate classes (Mayo 1 and Mayo 2) are increased significantly for all models. Due to high cost given to farther mispredictions, CDW-CE centers the wrong estimates mostly in classes with one neighborhood distance. Since mispredictions are more close to true classes in CDW-CE, we observe an increase in remission and non-remission sensitivities for the remission classification.

Figure \ref{fig:power_analysis} reveals that different models may have different optimum $\alpha$ parameters. While the performance increases as the $\alpha$ increases and gets to the optimum value, the model accuracy decreases sharply beyond it. As the $\alpha$ is an exponential term, increasing it beyond the optimum value results in high cost making the training unstable, resulting in an increase in standard deviation of cross-validation results (Figure \ref{fig:power_analysis}). Power analysis shows that a relatively high penalty given to distant classes (that can be counterintuitive at first) allows better optimization of the model training (for $\alpha=5$,  2-level neighborhood coefficient $(2^5)=32$ and 3-level neighborhood coefficient $(3^5)=243$). Nevertheless, $\alpha$ is not a very sensitive parameter for performance as Figure \ref{fig:power_analysis} shows that even the training with non-optimum $\alpha$ values outperforms the baseline and other ordinal approaches.

The experimental results show that using a loss function that penalizes distant mispredictions provides better optimization compared to previous approaches. While CDW-CE penalizes the mispredictions according to their distance to true class, it does not restrict the network to employ a single node at the output layer as it is in metric learning or regression-based approaches. Moreover, CDW-CE does not enforce fixed distances between classes and does not enforce any parametric distribution. Experiments show that, for the given problem where there are four distinct classes, the optimum $\alpha$ value is around six. We speculate that the $\alpha$ is susceptible to dataset, number of classes, and the employed model architecture; therefore, we recommend trying at least a few different values when deciding it. As Figure \ref{fig:power_analysis} shows, as the $\alpha$ increases, network training becomes unstable (i.e., cross-validation results vary a lot), so it is possible to get a high performance randomly from a single training. To avoid this trap, it is necessary to use methods such as cross-validation or multiple training with different seeds when deciding the $\alpha$.

\begin{figure}[t!]
  \centering
  \includegraphics[width=250pt, keepaspectratio]{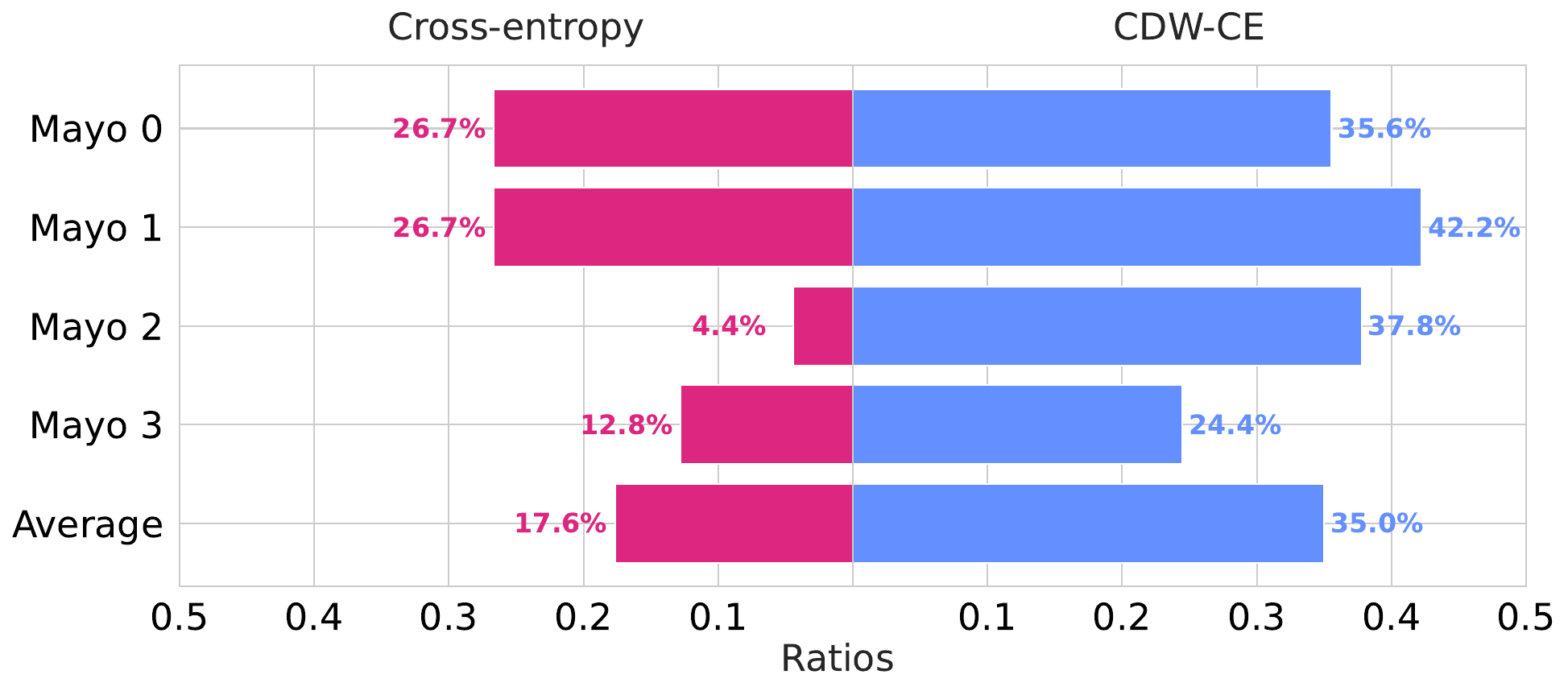}
  \caption{Assessment results of CAM visualizations of models trained with CE and CDW-CE by experts. The percentage values experts found both visualizations equally reasonable are as follows: 37.7\%, 31.1\%, 57.8\%, 62.8\%, 47.4\%, respectively.}
  \label{fig:cam_comparsion_result}
\end{figure}

Training models with the proposed CDW-CE loss does not only improve performance but also provide better explainability through the CAM visualizations. The model trained with CDW-CE highlights more relevant and discriminative regions compared to the model trained with CE for all Mayo scores. Sample CAM visualizations in Figure \ref{fig:cam_examples} show that using CDW-CE loss trains the model to extract more compatible features with the disease symptoms, leading to better performance. The CAM regions extracted by CDW-CE generally appear to be wider; however, these expansions were towards relevant regions rather than unrelated regions. Therefore, it can be said that CDW-CE has semantically captured better features. The average of the three experts' choices is shown in Figure \ref{fig:cam_comparsion_result}. The experts found that the CAM visualizations of the model trained with CDW-CE are more reasonable than the model trained with CE for all Mayo classes. On average, the experts found nearly half of the images equally reasonable (47.4\%) and the rate of selecting CDW-CE is two times more than the Cross-entropy (35.0\% vs. 17.6\%). Providing more reasonable CAM compatible with disease symptoms along with the high estimation performance increases the trust for the usage of the computer-aided diagnosis systems in clinics. As CDW-CE increases interpretability, transitioning to the clinic will also be accelerated.

MES for UC consists four distinct classes; therefore, experiments performed in this work are only compared for four levels. To what extend CDW-CE loss performs well should be investigated on different datasets, such as cervical cancer (7 levels of diagnosis) or diabetic retinopathy (5 levels of diagnosis) analysis. To test its capability in problems with higher number of classes, non-medical datasets, such as age estimation from face images, can be used. In addition, although the compared ordinal regression approaches are the state-of-the-art, other approaches based on regression setting can be experimented to extend the work.

\section{Conclusion}

In this study, we have proposed a novel non-parametric loss function designed to penalize the incorrect class predictions for the UC endoscopic severity estimation task. Incorrect classifications are weighted with a term that is in relation to its distance to the true class. Results show that a high penalty to the mispredicted distant classes is very important as experiments show that the optimal $\alpha$ can be a relatively large number. Extensive experiments show that the proposed loss function improves the performance significantly compared to the commonly used cross-entropy and several ordinal regression approaches. Training with CDW-CE does not only provide higher performance but also the models' CAM visualizations are more aligned with the experts opinions, which is expected to contribute positively to their clinical adoption. The proposed approach can be adapted to any problem with an ordinal category structure in medical as well as non-medical applications. 
In the future, we are planning to investigate its use in other ordinal regression problems.


\bibliographystyle{splncs04}
\bibliography{references}

\end{document}